# Climate-mediated shifts in temperature fluctuations promote extinction risk

Kate Duffy[1,2,3]*, Tarik C. Gouhier[4], and Auroop R. Ganguly[1,5]

[1] Sustainability and Data Sciences Laboratory, Department of Civil and Environmental Engineering, Northeastern University; Boston, Massachusetts 02115, USA.

[2] NASA Ames Research Center, Moffett Field, California 94035, USA.

[3] Bay Area Environmental Research Institute; Moffett Field, California 94035, USA

[4] Department of Marine and Environmental Sciences, Marine Science Center, Northeastern University; Nahant, Massachusetts 01908, USA

[5] Pacific Northwest National Laboratory, Richland, WA 99354

**\*Corresponding author email: duffy.m.kate@gmail.com**

**Abstract**

Climate-mediated changes in thermal stress can destabilize animal populations and promote extinction risk. However, risk assessments often focus on changes in mean temperatures and thus ignore the role of temporal variability or structure. Using Earth System Model projections, we show that significant regional differences in the statistical distribution of temperature will emerge over time and give rise to shifts in the mean, variability, and persistence of thermal stress. Integrating these trends into mathematical models that simulate the dynamical and cumulative effects of thermal stress on the performance of 38 globally-distributed ectotherm species revealed complex regional changes in population stability over the 21st century, with temperate species facing higher risk. Yet despite their idiosyncratic effects on stability, projected temperatures universally increased extinction risk. Overall, these results show that the effects of climate change may be more extensive than previously predicted based on the statistical relationship between biological performance and average temperature.

**Main**

Biodiversity loss has been recognized as one of the top global risks by the World Economic Forum because it could erode or eliminate key ecosystem functions and services[1]. Climate change is expected to surpass habitat loss as the leading threat to global biodiversity by the middle of the 21st century[2]. Observed changes in the distribution and phenology of species have already been linked to climate fluctuations in numerous studies[3]. Although conservation actions may ameliorate potential biodiversity loss, the success of these efforts depends on our ability to predict the response of ecological systems to environmental changes.

Most ecological impact studies to date have relied on statistical models such as bioclimate envelope approaches to determine how climate change will impact ecological populations[4–7]. Bioclimate envelope models are typically constructed by either mapping the geographical distribution of species to co-located temperature records via regression techniques or by building species' thermal profiles via empirical assessments of their performance across a range of temperatures (i.e., thermal performance curves or TPCs)[4,8]. These relationships between organisms and temperature are then used to predict the distribution of species under future thermal conditions projected under various climate change scenarios.

Despite the power and popularity of TPCs, these statistical approaches can yield inaccurate predictions because they typically rely on mean annual conditions and thus ignore the influence of the temporal structure of temperature fluctuations at finer scales. This is problematic because the nonlinear relationship between temperature and most metrics of biological performance essentially guarantees that the average organismal response will not be equivalent to their response to the average condition[9–12]. Specifically, when an organism is exposed to a sequence of temperatures $x$, its performance at the average temperature $f(\bar{x})$ will differ from the average of its performance $\overline{f(x)}$. Temporal variation in temperature will either magnify $\left(\overline{f(x)} > f(\bar{x})\right)$ or dampen $\left(\overline{f(x)} < f(\bar{x})\right)$ the effects of its mean on organismal performance depending on the curvature of $f$ (i.e., whether $f$ is accelerating or decelerating[9]). In many cases, changes in temperature variability can be as or more important than changes in the mean value[13,14]. In one study, climate-mediated changes in mean temperature alone were found to broadly promote organismal performance in ectotherms, but accounting for the temporal variability of temperature dampened this effect and led to most species suffering a performance loss[15].

Although the temporal structure of temperature can theoretically be incorporated into bioclimate envelope models by using finer temporal scale data, accounting for its dynamical effects on organisms is much more difficult because of the 'static' nature of these methods and their general inability to account for the cumulative effects of previous temperatures on organismal performance. However, theory has shown that such carry-over effects associated with the temporal structure or autocorrelation of temperature can interact with the magnitude of temperature variability to determine population persistence[16]. Specifically, temporally autocorrelated variation tends to reduce extinction risk by decreasing the likelihood of catastrophic conditions under strong variation, whereas temporally autocorrelated variation tends to promote extinction risk under weak variation by increasing the likelihood that organisms will experience long stretches of poor conditions[16]. Prolonged exposure to temperatures above the species critical thermal maximum is particularly destabilizing as it can reduce population fitness below the replacement rate[17]. Analyses of historical observations and projections from previous generation climate models have found strong temporal trends in the variability and autocorrelation of temperature[18–21], suggesting the potential for a larger impact on ecological populations in the future. Overall, these empirical and theoretical results highlight the importance of quantifying changes in the mean, variability, and autocorrelation of temperature projected under climate change to predict their joint influence on ecological systems over the course of the 21st century. However, disparities in the scale of models in climate and ecology have hindered impact studies that consider the complexity of both underlying systems[22,23].

We briefly illustrate the potential for complex interactions between climate-mediated changes in the mean, variability, and autocorrelation of temperature to influence organismal performance by simulating the effects of synthetic temperature time series on the population growth rate $r$ according to a species' TPC (Fig. 1, see Methods for modeling details). Predictably, performance under negligible temperature variation can be inferred directly from the mean of each species' TPC (Fig. 1b,c). However, when temporal variation in temperature is included in the model (i.e., standard deviation; shaded region), time-averaged performance can be considerably modified[9], even overturning the identity of 'winning' and 'losing' species based solely on constant temperature conditions (Fig. 1d,e). Temperature autocorrelation, which measures the temporal structure of temperature fluctuations (e.g., the persistence of extremes), can also play a pivotal role in determining whether a species' performance and stability will benefit or suffer under different thermal regimes (Fig. 1f,g). To determine the impact of such changes over the course of the 21st century, we analyzed the latest generation of Earth System Models from the Coupled Model Intercomparison Project Phase 6 (CMIP6) in order to document spatiotemporal changes in three key aspects of air temperature: statistical distribution, variance, and temporal autocorrelation. We then analyzed the effects on population stability and extinction risk using simple mathematical models to examine the hypothesis that even under ideal conditions, popular statistical methods can yield incorrect predictions about patterns of organismal performance when dynamical and cumulative temperature effects are ignored.

**Regional trends in temperature distribution**

We examined changes in the global and regional temperature distributions at each geographical location between 1850 and 2100 under the high emissions scenario, SSP5-8.5[24] (Fig. 2a,b). Quantile regression was used to measure temporal trends in the entire distribution of projected temperatures (i.e., across quantiles ranging from $\tau$ = 2.5% at the low end to $\tau$ = 97.5% at the high end) in the Northern Hemisphere Extra-tropics (NHEX; 30°N to 90°N), the Southern Hemisphere Extra-tropics (SHEX; 90°S to 30°S), and the Tropics (TROP; 30°S to 30°N). When averaging trends across regions, we found asymmetrical but uniformly positive trends across all quantiles, indicating that the entire temperature distribution is shifting upwards, but at rates that vary systematically across the distribution. In NHEX, the lowest quantile of the distribution ($\tau$ = 2.5%, 0.33 K decade$^{-1}$) is warming at twice the rate of the uppermost quantile ($\tau$ = 97.5%, 0.16 K decade$^{-1}$). The SHEX exhibits a similar pattern of disproportionate warming for the low quantiles ($\tau$ = 2.5%, 0.15 K decade$^{-1}$; $\tau$ = 97.5%, 0.10 K decade$^{-1}$). Conversely, in the tropics, the upper quantiles of temperature are warming faster ($\tau$ = 97.5%, 0.14 K decade$^{-1}$) than the lower quantiles ($\tau$ = 2.5%, 0.10 K decade$^{-1}$). The magnitude of trends is greater in NHEX than in SHEX or TROP. The more pronounced extra-tropical decrease in the incidence of cold events may benefit cold-limited species, however, quantile trends also indicate increased positive skewness of the NHEX temperature distribution, which has been associated with declines in long term ecological performance[15]. Across all eight CMIP6 models that we analyzed and in all three latitudinal regions, trends in the tails of the distributions differ from the trends in the central tendencies, thus highlighting the importance of moving beyond mean temperature when predicting organismal performance.

Trends in the variability of temperature between 1850 and 2100 are predicted to exhibit similarly complex regional patterns (Fig. 2c). Variance will generally increase across temperate and tropical land areas below 45°N, with regional exceptions including Asia. The strongest increases in variance are in the northern midlatitudes, including northern Africa, southern Europe, the Middle East, and the western United States. Variance is decreasing most rapidly in the high northern latitudes, especially in Canada and Russia[25]. The concurrent decrease of variability at high latitudes and its increase at other latitudes suggests that temperature variation, like mean temperature, is becoming more spatially homogeneous in a warming world. These findings are generally consistent with studies of the previous generation of climate models, which suggested increasing temperature variability in tropical countries[26] and decreasing variability in the northern mid- to high- latitudes[27]. Trends at the regional level are congruent with quantile trends (Fig. 2a), which indicate a widening temperature distribution (increasing variance) in TROP, and a narrowing temperature distribution (decreasing variance) in NHEX and SHEX, as well as large scale changes in physical climate processes[26–28]. The effects of these trends in temperature variation on ecological systems will depend on the geographical location and physiological properties of each species, with increasing variability either promoting or reducing performance based on its position relative to the inflection point of an organism's TPC[9].

**Frequency-resolved temperature changes**

To better understand these spatiotemporal patterns, we used time-frequency decomposition via the wavelet transform to resolve changes in the variability of temperature at sub-annual to annual timescales (between 2 days and 2 years) and multiannual timescales (between 2 years and 30 years; Extended Data Fig. 1). Wavelet transforms resolve a signal in both the time and frequency domains to describe how each frequency or period in the time series contributes to variation over time. We found countervailing trends in scale-specific variability in the mid to high-northern latitudes. The magnitude of short-term variability is decreasing, while the magnitude of long-term variability is increasing. Arctic amplification, which is detectable in both observational data and climate simulations, has previously been suggested as the main driver of decreasing sub-seasonal variability at these latitudes[27]. Meanwhile at the mid latitudes, variation at both annual and multiannual time scales is increasing, consistent with increasing variance at all periodicities. These scale-dependent changes in the temporal trends of temperature fluctuations could have important ecological implications because the effect of temperature fluctuations depends on the relationship between their period and the generation time of organisms. Indeed, estimating the biological effect of temperature fluctuations by 'nonlinear averaging' organismal performance under the relevant constant thermal regimes is much more likely to yield accurate results when the period of the temperature fluctuations is larger than the generation time of an organism because such slow variation can more easily be "tracked" by a population[29].

We computed the spectral exponent of the temperature time series at each geographical location to quantify spatiotemporal trends, with more negative exponents indicating greater temporal autocorrelation over a range of lags from 2 days to 10 years (Fig. 3a). We found increasing temporal autocorrelation (decreasing spectral exponent) at a majority of sea locations (60%) and land locations (80%), excluding Antarctica where autocorrelation is decreasing. Autocorrelation is increasing most rapidly in equatorial land areas including the Amazon and the Southeast Asian islands with high inter-model agreement on the sign of the trend. Notable exceptions to the increasing trend in autocorrelation include Greenland, Western Africa, Western Europe, and parts of Central Asia. Generally, agreement between models is higher at mid-latitudes than in the polar zones or the tropics, where climate model bias and spread have historically persisted[30]. Regional analysis indicates statistically significant increasing trends in temporal autocorrelation in NHEX (-1.12e$^{-3}$ decade$^{-1}$, p-value=0.010), TROP (-1.14e$^{-3}$ decade$^{-1}$, p-value=0.001), and globally (-0.54e$^{-3}$ decade[-1], p-value=0.005), and a statistically significant decreasing trend in temporal autocorrelation in SHEX (0.53e[-3] decade[-1], p-value=0.009; Supplementary Table 1). The direction and significance of these trends are consistent across land and sea environments, although the spectral exponent is more negative for sea than land, likely due to the buffering effects of the ocean (Fig. 3b-e). In NHEX and TROP autocorrelation is increasing at a greater rate in land locations than sea locations while in SHEX autocorrelation is decreasing at similar rates between land and sea (Supplementary Table 2). A greater degree of temporal autocorrelation is associated with more gradual changes of state, and, even absent any changes in variance, results in longer durations spent under extreme conditions. A greater clustering of similar temperatures has been suggested to increase exposure to heat waves and cold snaps while decreasing the incidence of protective temporal refugia[20].

**Regional differences in warming patterns**

In the northern latitudes, variance and autocorrelation exhibit opposite temporal trends. The decreasing variance may be attributed to a decrease in high frequency variability and more rapid warming of the lower than upper quantiles of the temperature distribution. Studies of reanalysis data and observations have also implicated decreasing cold-season sub-seasonal variability and rapidly warming cold days in decreasing temperature variability in mid to high northern latitudes[20,24,29]. Meanwhile, temporal autocorrelation in NHEX is increasing, a finding which has also been detected in the previous generation of climate models[20], weather station observations[32], and monthly reanalysis data[19]. As a result, variation at 2-day to 10-year periodicities is decreasing while temperature fluctuations are becoming more persistent, suggesting the increased probability of a series of homogeneous conditions. In contrast to the mid to high northern latitudes, variance and temporal autocorrelation show similar trends at most latitudes, that is, both variance and autocorrelation are increasing.

**Implications for global ectotherm populations**

To better understand the independent and joint effects of these projected trends in the mean, variance, and autocorrelation of temperature on ecological systems, we used empirical thermal performance information from invertebrate ectotherms compiled by Deutsch et al. (2008). We extracted temperature time series from the eight CMIP6 climate models at geographical point locations corresponding to the source sites of the 38 species (Fig. 4a). A dynamical population simulation using species-specific temperature-dependent growth rates yielded time series of population abundance for the historical period (1950-2000) and the latter half of the 21st century (2050-2100). We used a dynamical logistic growth model whose carrying capacity $K = r_t/\alpha$ was determined by the temperature-dependent growth rate $r_t$ and the self-regulation parameter $\alpha$. Importantly, the model captures the effects of temperatures above the critical thermal maximum and extinction propensity under autocorrelated variation by allowing growth rates to become negative (see Methods for details). Using the eight climate simulations as replicates, we compared the historical and future periods to detect statistically significant temperature-driven changes in population abundance, stability (mean/standard deviation of abundance), and extinction probability (proportion of simulations where a species did not have a strictly positive final abundance).

Under the high emissions scenario (SSP5-8.5), population abundance increased for the plurality of species (18 of 38) because the mean temperature grew closer to their thermal optimum and thus boosted equilibrium abundance, but it decreased for 10 species (Supplementary Table 3). Population abundance increased significantly for all TROP species (5 of 5) and for the majority (5 of 8) of SHEX species. In NHEX, outcomes were mixed with approximately equal proportions of species experiencing an increase in abundance, a decrease in abundance, and no significant change. NHEX population abundance followed latitudinal patterns, generally decreasing between 30°N and 45°N, and increasing above 45°N. Under the high emissions scenario, population stability increased for the plurality of species (16 of 38) and decreased for 10 species (Fig. 4b). Population stability increased or underwent no significant change for TROP species, while in the mid-latitudes (NHEX and SHEX), changes in stability were mixed. Additional analyses showed that the trends in stability were mainly due to the emergence of two distinct dynamical regimes under climate change, with species either moving to a low-mean/low-variance mode or a high-mean/high-variance mode, particularly in the extra-tropics (Extended Data Fig. 2-3). These results were robust to orders of magnitude changes in the growth rate $r_t$ and self-regulation parameter $\alpha$ (Extended Data Fig. 4-5).

Many SHEX and NHEX species suffered performance losses (negative growth rates) during summers in their respective hemispheres, as they are generally less tolerant of hot temperatures than tropical species. For some temperate species, longer growing seasons and warmer winter temperatures offset the negative effect of the warmest part of the year, while others will suffer an overall performance loss[33]. This is consistent with the suggestion that increases in summer heat stress would reduce overall fitness and increase fitness variation for many mid-latitude species . Our results suggest that temperate species may be at greater risk than tropical species as a result of warm days, even when annual mean temperature remains below the thermal optimum. The results contrast with those of previous studies, which suggested based on hourly temperature records and monthly temperature anomalies that warming in the tropics would be more deleterious than warming in the mid-latitudes[5,34]. This discrepancy may be due to the fact that growth rates were allowed to become negative when temperatures exceeded the critical thermal maximum in our simulations but assumed to converge to zero (i.e., were not allowed to be negative) in previous studies[4]. Our results are more consistent with studies that predict a greater risk of performance loss for temperate species when accounting for negative performance values in response to climate-mediated changes in the mean and the variance of temperature[15].

To tease apart the dynamical effects of climate change on population stability from its effects on mean performance as inferred by measuring average growth rate using each species' TPC, we replicated previous efforts by comparing changes in the average growth rate under historical and future climatic conditions with vs. without negative growth rates (Extended Data Fig. 6). Our results show that although allowing negative growth rates predictably leads to greater reductions in performance overall, the regional patterns in performance are similar to the trends in population stability observed in the dynamical simulations, with tropical species generally enjoying performance gains and temperate species—particularly in NHEX—suffering performance losses (Extended Data Fig. 6).

Our simulations indicated mean warming as the dominant driver of ecological impacts. Changes in temporal autocorrelation alone (mean temperature and variance held at historical levels) had no significant effects on population abundance and a significant destabilizing effect on just 3 NHEX species. Changes in temporal autocorrelation and variance (mean temperature held at historical levels) led to a decrease in population abundance in 2 NHEX species and a decrease in population stability in 5 NHEX species. These results suggest that NHEX species are more vulnerable to negative effects of changes in temperature variability than TROP or SHEX species. Finally, changes in mean and temporal autocorrelation (variance held at historical levels) led to increased population abundance in 19 global species and increased stability in 19 global species, versus 18 and 16 under the high emissions scenario projected changes in all three aspects of temperature. Thus, projected changes in temperature variability have a weak moderating effect on the positive effects of mean warming on population abundance and stability.

To determine how these complex changes in population abundance and stability translate to persistence, we quantified extinction risk as the proportion of the eight CMIP6 models for which population abundance declined below an arbitrarily small threshold of 1e-9 at any point during the 50-year simulation (Fig. 4c). In our simulations under the high emissions scenario, extinction risk increased significantly under future climate conditions relative to historical baselines for 25 species, increased (but not significantly) for 13 species, and decreased for 0 species. We found statistically significant increases in extinction risk globally (Mann–Whitney $U = 376$, $n_1 = n_2 = 38$, p-value = 6e-5) and in NHEX (Mann–Whitney $U = 150.5$, $n_1 = n_2 = 25$, p-value=5e-4). These findings suggest that temperature changes promote extinction risk, despite having a largely positive or neutral effect on population abundance and idiosyncratic impacts on stability. Hence, although variability among climate models produces a wide range of changes in stability across species and geographical locations, uncertainty at the climate level yields consistent biological impacts in the form of systematically higher extinction risks (Extended Data Fig. 7).

**Conclusion**

By forcing simple strategic and dynamical models of population growth with fine temporal scale temperature projections from the latest generation of Earth System Models, we demonstrated increased extinction risk under climate change across globally-distributed ectotherm populations. Unfortunately, using more complex tactical dynamical models would require extensive species-, age-, and life-stage specific information about the effects of temperature fluctuations on population growth rates that is simply not available at the relevant scales. Tactical models would also need to consider thermoregulation[35], the effects of microclimates[36], acclimatization or adaptation[37], partitioning of activity periods[38], and synecological processes such as predator-prey interactions that could affect ectotherm population dynamics. Additionally, due to their 1° spatial resolution, the climate projections used in this study are much coarser than the microclimates experienced by individual organisms and may thus lead to underestimates of organismal performance due to the presence of thermal refugia in the real world[23,35]. Hence, our results should be viewed as a qualitative baseline prediction of how the spatiotemporal distribution of extinction risk is likely to shift due to climate change rather than a quantitative forecast of when each species is likely to be extirpated from each geographical location.

Despite the limitations of TPCs in accounting for temporal carryover and dynamical effects, the lack of obvious alternatives calls for strategies to make these approaches more robust to real-world conditions[39], such as by integrating more realistic, fine-scaled temperature variation into our predictive models than previous studies. Although bioclimate envelope approaches have been criticized for not accounting for important ecological factors such as species interactions and dispersal when attempting to predict the ecological effects of climate change[40–43], we have shown that even under ideal conditions when the influence of such factors can be assumed to be negligible, statistical frameworks that ignore the dynamical consequences of temperature variation are likely to yield inaccurate forecasts of the impact of climate change on organisms. Our results show that accounting for shifts in the entire statistical distribution of temperature over time via dynamical models can better capture the cumulative effects of climate-mediated changes in thermal stress on extinction risk.

By bringing together climate data and a minimal dynamical model from ecology, we demonstrated a strong and systematic amplification of extinction risk in ectotherms due to projected changes in fine-grained temperature variability. Furthermore, our finding of greater risk to sub-tropical than tropical species highlights the importance of accounting for the dynamical effects of projected changes in the mean as well as variance of temperature over the course of the 21st century to accurately predict the response of ecological systems around the globe.

**Acknowledgments**

This work was primarily supported by the National Science Foundation (NSF) grant CCF-1442728 while KD was a PhD student at the SDS Lab in Northeastern University. Furthermore, additional support was provided for KD and ARG by NSF SES-1735505 and for TG by NSF OCE-2048894. The authors gratefully acknowledge the background support from a prior NSF Expeditions in Computing grant (award # 1029711) and an ongoing DOD Strategic Environmental Research and Development Program funding (# RC20-1183). KD and ARG acknowledge support from the NASA Ames Research Center.

**Author Contributions Statement**

KD, TG and ARG conceived, designed and refined the project, KD performed the data analysis and modeling, KD, TG and ARG interpreted the results, and KD wrote the manuscript with contributions from TG and ARG.

**Competing Interests Statement**

The authors declare no competing interests.

**Methods**

CMIP6 simulations

We obtained CMIP6 climate simulations for the historical forcing period (1850-2014) and future emissions scenario SSP5-8.5 (2015-2100) via the CMIP6 data portal (https://esgf-node.llnl.gov/search/cmip6/). Eight models from CMIP6 (AWI-CM-1-1-MR, BCC-CSM2-MR, CESM2, EC-Earth3, INM-CM5-0, MPI-ESM1-2-HR, MRI-ESM2-0, and NorESM2-MM) were selected based on availability of daily air temperature at surface ("tas") at a 100 km nominal resolution at the time of download. While "tas" at sub-daily frequencies is available for some models, daily data was selected to maximize the ensemble size. We resampled all datasets to a common 1° by 1° grid spanning -90° to 90° latitude and 0° to 360° longitude, and to a standard calendar without leap years. Spatial regions were defined based on latitude as Northern Hemisphere Extra-tropics, 90°S to 30°S; Tropics, 30°S to 30°N; and Southern Hemisphere Extra-tropics, 30°N to 90°N.

**Statistical analyses of climate data**

Quantile regression

Trends in the percentile values global and regional temperature distributions were computed via quantile regression. Quantile regression can comprehensively model heterogenous conditional distributions, where the relationship between the quantiles of the dependent variable and the independent variable is different from the relationship between the mean of the dependent variable and the independent variable. We applied quantile regression to analyze trends with respect to time at various percentile values ($P_{2.5}$, $P_{10}$, $P_{20}$, $P_{30}$, $P_{40}$, $P_{50}$, $P_{60}$, $P_{70}$, $P_{80}$, $P_{90}$, $P_{97.5}$). Analyses were performed using the R package quantreg, with significance level $\alpha = 0.1$ and the default Barrodale and Roberts method to return confidence intervals for the estimated parameters. To obtain the ensemble mean trends, we calculated the mean slope, upper bound, and lower bound across the eight climate models at each geographical location, then computed spatial averages for the full globe and three latitudinal regions.

Variance

Trends in the magnitude of temporal variation of air temperature were examined at each geographical location using a moving window approach. First, temperature was detrended by fitting a piecewise linear regression against time with Python package pwlf at each geographical location and extracting the residuals. Then, the temperature time series were divided into 10-year windows starting in years 1855 through 2085 so as not to combine historical and future simulations (pre- and post- 2015-01-01), and the variance of daily air temperature was calculated for each window. Windows were selected with no overlap to avoid statistical issues due to non-independence of estimates taken from partially overlapping time windows[20].

Scale-specific variability

Scale-specific variability was quantified using time-frequency decomposition. Specifically, at each geographical location, wavelet analysis was conducted on multi-model mean temperature using the R package biwavelet[44]. Wavelet analysis resolves both the time and frequency domains of a signal (here a time series) via the wavelet transform. This is achieved via the convolution of a mother wavelet function and a time series across a set of windows $\tau$ and scales $s$. We chose to the Morlet wavelet, which represents a sine wave modulated by a Gaussian function[45]:

$$\psi_0(t) = \pi^{-1/4} e^{i\omega_0 t} e^{-r^2/2}$$

Where $i$ is the imaginary unit, $t$ represents nondimensional time, and $\omega_0 = 6$ is the nondimensional frequency[3]. The continuous wavelet transform of a discrete time series $x(t)$ with equal spacing $\delta t$ and length $T$ is defined as the convolution of $x(t)$ with a normalized Morlet wavelet[45,46]:

$$W_x(s,\tau) = \sqrt{\frac{\delta t}{s} \sum_{t=0}^{T-1} x(t)\psi_0 * \left(\frac{(t-\tau)\delta t}{s}\right)}$$

where $*$ indicates the complex conjugate. By varying the wavelet scale $s$ (i.e., dilating and contracting the wavelet) and translating along localized time position $\tau$, one can calculate the wavelet coefficients $W_x(s,\tau)$ across the different scales $s$ and positions $\tau$. These wavelet coefficients can be used to compute the bias-corrected local wavelet power, which describes how the contribution of each frequency or period in the time series varies over time[45,47,48]:

$$W_x^2(s,\tau) = 2^s |W_x(s,\tau)|^2$$

Where $2^s$ is the bias correction factor[47]. The scale $s$ of the Morlet wavelet is related to the Fourier frequency $f$ [48,49]:

$$\frac{1}{f} = \frac{4\pi s}{\omega_0 + \sqrt{2 + \omega_0^2}}$$

When $\omega_0 = 6$, the scale $s$ is approximately equal to the reciprocal of the Fourier frequency $f$ so period $p \approx s$. The local wavelet power spectrum can then be visualized via heatmaps and contour plots[46,48]. From the resulting local wavelet power spectrum heatmap with time on the x-axis, period (scale) on the y-axis, and power on the z-axis, scale-averaged wavelet power was computed at annual (between 3 days and 2 years) and multiannual (between 2 years and 30 years) periodicities. This was achieved by taking the weighted sum of the local wavelet power across all scales for each time location $\tau$ [45,48]:

$$W_x^2(\tau) = \frac{\delta j \delta t}{C_\delta} \sum_{j=0}^{J} \frac{|W_x(s_j,\tau)|^2}{s_j}$$

where $C_\delta = 0.776$ for the Morlet wavelet, $\delta J$ represents the spacing between successive scales and $\delta t$ represents the spacing between successive time locations[45]. Scale-averaged power was then regressed against time using Generalized Least Squares (GLS) regression for the period of 1850-2100 at each geographic location. To determine the robustness of results to the choice of period for scale averaging, we also performed analysis of trends separately at interannual (between 2 years and 7 years) and multiannual (between 7 years and 30 years) scales and found qualitatively similar results.

Temporal autocorrelation

The temporal autocorrelation of air temperature was quantified by calculating the spectral exponent at each geographical location[20]. As described above, temperature was detrended by fitting a piecewise linear regression at each geographical location and extracting the residuals.

The detrended temperature was divided into 10-year windows starting in years 1855 through 2085. Fourier transforms of each time series were computed via fast Fourier transform using the Python package NumPy. Periodograms were prepared with frequency on the x-axis and power spectral density on the y-axis. The spectral exponent, β, was calculated as the slope of the regression line relating log transformed power to log transformed frequency. β expresses the relative contributions of frequencies to the power spectrum. In the case of equal contribution from all frequencies, β = 0. Greater contribution from low frequencies than high frequencies results in a more negative value of β, and indicates greater temporal autocorrelation in the time domain.

<u>Analysis of decadal trends</u>

For each climate model, Generalized Least Squares (GLS) regression was used to detect statistically significant trends (p-value < 0.05) in variance and temporal autocorrelation with respect to time in the presence of potentially autocorrelated residuals. To measure inter-model agreement, we calculated the multi-model mean trend as the mean of trends calculated for each of the 8 models at each geographic location, then assessed the proportion of models that agreed with the sign of the multi-model mean trend. Inter-model agreement was considered as statistically significant at the $\alpha = 0.1$ level based on a binomial test. ANCOVA was used to quantify the relationship between temporal autocorrelation and time while accounting for potential differences between land and sea environments. Statistically significant main effects and interactions were reported for p-value < 0.05.

**Modeling temperature impacts on ecology**

<u>Thermal tolerance data</u>

We obtained experimentally derived thermal tolerance parameters for a set of terrestrial ectotherms (n = 38) published by Deutsch et al. (2008) and used them to predict physiological response to CMIP6 simulated temperature. Deutsch et al. gathered data from 31 thermal performance studies published between 1974 and 2003 based on a collection of insects from 35 different locations. For each species, experimental intrinsic growth rates at multiple temperatures were used to fit a TPC yielding least-squares estimates of key parameters such as critical thermal maximum ($CT_{max}$), optimum temperature ($T_{opt}$), and sigma ($\sigma$). We used a numerical scheme to reconstruct the curves whereby the rise in performance up to $T_{opt}$ was modeled as Gaussian and the decline beyond $T_{opt}$ was quadratic[5,50]

$$P(T) = \begin{cases} \exp\left[-\left(\frac{T-T_{opt}}{2\sigma}\right)^2\right] & \text{for } T \leq T_{opt} \\ 1-\left(\frac{T-T_{opt}}{T_{opt}-CT_{max}}\right)^2 & \text{for } T > T_{opt} \end{cases} \quad [1]$$

This allowed negative growth rates to arise at high temperatures but growth rates were bound at zero at low temperatures. Negative performance values indicate that mortality surpasses reproduction rates. Because $P(T)$ is capped at 1 under this numerical scheme, $P(T)$ represents the relative fitness of each species based on its normalized maximum growth rate. However, scaling this relative or normalized maximum growth rate by two orders of magnitude (i.e., by a factor of 0.1 or 10.0) had limited quantitative and no qualitative impact on our results (Extended Data Fig. 4). Overall, increasing the growth rate scaling factor had no impact on population stability but promoted extinction risk.

<u>Isolation of temperature aspects</u>

To isolate projected changes in mean temperature and variability, we transformed the future (2050-2100) time series using z-score normalization. Using this approach, we modified projected time series to match the historical (1950-2000) mean and/or standard deviation. Working in 10 year moving windows between 2050 and 2100, each series $x_i$ with mean $m_1$ and standard deviation $s_1$ was transformed to series $y_i$ with mean $m_2$ and standard deviation $s_2$:

$$y_i = m_2 + (x_i - m_1)\frac{s_2}{s_1} \quad [2]$$

According to the scenario, $m_2$ and $s_2$ were alternatively defined as [1] high emissions scenario mean and standard deviation ("Mean, variance, and autocorrelation"), [2] high emissions scenario mean and historical standard deviation ("Mean and autocorrelation"), [3] historical mean and high emissions scenario standard deviation ("Variance and autocorrelation"), and [4] historical mean and standard deviation ("Autocorrelation"). High emissions scenario statistics refer to the properties of future series $x_i$ and confer no change to that aspect of the time series.

Population dynamical modeling

To model the effects of temperature change on the stability and extinction probability of global ectotherm populations, temperature dependence was integrated in the growth rate term of a population dynamical model[51]. While more complex synecological models can capture a range of community-level effects including competition and predation, we chose to model first order autecological dynamics in order to produce foundational insights about the role of temperature fluctuations on single-species population dynamics. Specifically, we used the $r - \alpha$ logistic growth model to simulate temperature-dependent growth and negative density-dependence:

$$\frac{dN}{dt} = N(r_t - \alpha N) \qquad [3]$$

with population size $N$, time $t$, temperature-dependent growth rate $r_t$, and self-regulation in the form of intraspecific competition $\alpha$. This $r - \alpha$ logistic model is easily interconvertible with the classical $r - K$ formulation ($r/\alpha = K$), but has the advantages of handling negative values of $r$ without issues[52]. This approach is sensitive to the effects of temperatures at and above the critical thermal maximum, which can yield negative growth rates that are important for determining population dynamics as well as long term fitness.

We extracted times series of daily temperature at the source locations for each species from the ensemble of eight climate simulations. Daily intrinsic growth rates were computed from temperature using Eqn. 1, incorporated into the $r - \alpha$ logistic growth model depicted in Eqn. 3, and the model was then numerically solved using the explicit Runge-Kutta method of order 5(4) implemented in the Python SciPy package in order to obtain daily population densities. Rather than delineating active periods, which may shift under climate change, we considered the full year to account for potential changes in fitness due to shifts in activity.

The sensitivity of the results to strong ($\alpha = 1$) and weak ($\alpha = 0.1$) self-regulation was examined and found to be extremely limited (Extended Data Fig. 5). We also assessed the sensitivity of our results to absolute rather than relative or normalized growth rates by scaling $r_t$ by a factor of 0.1 or 10 in our simulations. Scaling $r_t$ by two orders of magnitude in this manner had very little quantitative and no qualitive impact on our results. This suggests that the effects of temperature fluctuations on changes in the spatiotemporal distribution of population abundance, stability, and extinction were not contingent upon the use of relative fitness (i.e., normalized growth rate) versus absolute fitness (i.e., growth rate scaled by a factor of 0.1 or 10). These sensitivity analyses also served to show that our results are robust to temperature-mediated changes in the maximum instantaneous growth rate[53,54].

Analysis of population changes

To quantify temperature-driven changes in ecological stability and extinction probability, we compared population sizes and dynamics between a historical period (1950-2000) and a future period (2050-2100). Here, we defined latitudinal regions according to traditional delineations in ecology: Northern Hemisphere Extra-tropics, 60°S to 23°S; Tropics, 23°S to 23°N; and Southern Hemisphere Extra-tropics, 23°N to 60°N.

Population abundance was computed as the mean population size ($N$) for a time period. Population stability was computed as the inverse of the coefficient of variation, or mean population divided by population standard deviation. Percent changes in population size and stability were computed for each of the climate models as (future − historical)/historical × 100% and plotted without outliers in Fig. 4. Statistically significant changes in population abundance and stability between the historical and future periods were identified via the Mann-Whitney *U*-test with the eight models as replicates.

Extinction probability was quantified as the proportion of ensemble simulations for which the population declined to zero during a 50-year simulation. Changes in extinction probability were calculated as the difference between future and historical extinction probability. Statistically significant changes in extinction probability were identified on a regional basis via the Mann-Whitney *U*-test.

### Data availability

The CMIP6 simulation data used in this paper is available via the data portal https://esgf-node.llnl.gov/search/cmip6/. The ecology data is available for download at https://doi.org/10.1073/pnas.0709472105.

### Code availability

The code can be accessed on GitHub at https://github.com/KateDuffy/climate-change-ecology[55].

### Methods-only references

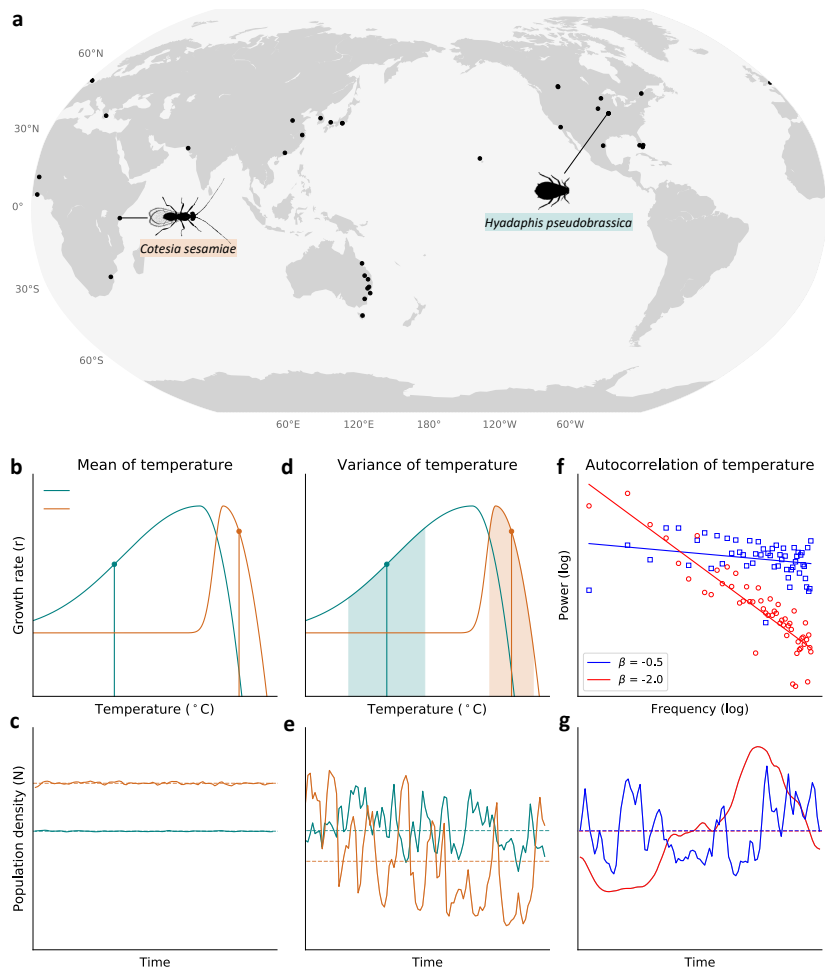

**Figure 1. Effects of temperature mean, variance, and autocorrelation on organismal performance**

**a**, Source locations of the 38 species whose thermal performance parameters were obtained from the Deutsch et al. (2008) dataset. *Cotesia sesamiae* is a tropical parasitoid wasp and *Hyadaphis pseudobrassicae* a temperate-zone turnip aphid. **b**, **c**, Thermal performance curves and population dynamics for *C. sesamiae* and *H. pseudobrassicae* under negligible temperature variation. **d**, **e**, Larger temperature variation (standard deviation shaded) alters mean response and may even overturn predictions of relative performance based on constant temperature conditions. **f**, The power spectrum of temperature with weak (ß=-0.5) and strong (ß=-2) temporal autocorrelation. **g**, Population dynamics of *Hyadaphis pseudobrassicae* under a greater degree of temporal autocorrelation exhibit longer-term fluctuations. Multiple aspects of temperature such as its mean and variance can interact to promote or decrease performance.

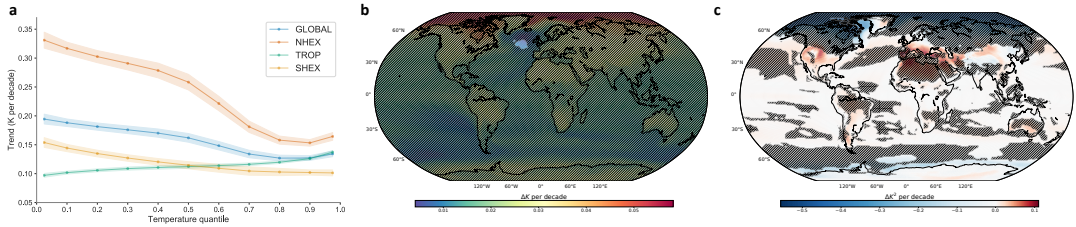

**Figure 2. Mean trends in the statistical distribution of daily air temperature between 1850 and 2100.**
Trends in the percentile values of air temperature (**a**; K/decade) and mean temperature at each geographic location (**b**; K/decade) indicate asymmetrically warming temperature distributions in the Northern Hemisphere Extra-tropics (NHEX; 30°N to 90°N), the Tropics (TROP; 30°S to 30°N), the Southern Hemisphere Extra-tropics (SHEX; 90°S to 30°S), and the full globe (GLOBAL; 90°S to 90°N). Shaded bounds denote a 90% confidence interval based on eight CMIP6 models. **c**, Trends in the variance of daily air temperature (K$^2$/decade) exhibit similarly complex regional patterns. The concurrent decrease of variability at high latitudes and increase at other latitudes suggests that temperature variation is becoming more spatially homogeneous in a warming world. Hashed contours indicate statistically significant inter-model agreement on the sign of the trend at the α = 0.05 significance level.

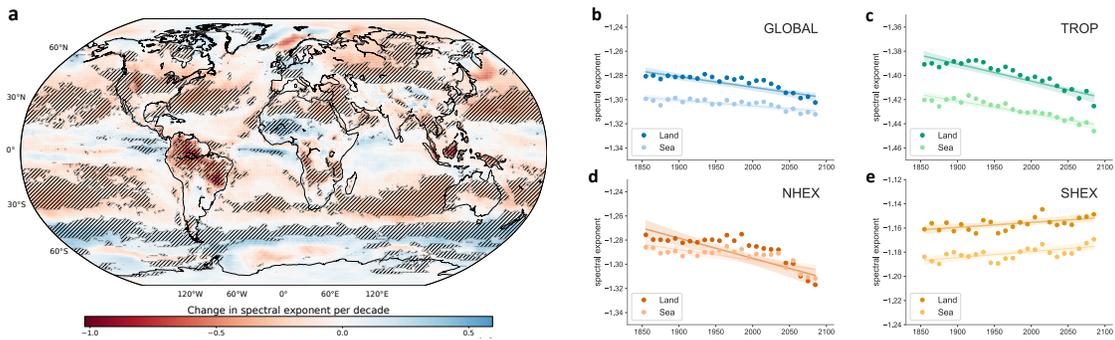

**Figure 3. Increasing temporal autocorrelation in daily air temperature between 1850 and 2100.**
**a**, Spatiotemporal trends in temporal autocorrelation suggest changes in the chronological sequence of temperature conditions, with increasing temporal autocorrelation (decreasing spectral exponent) at 80.04% of global land locations, excluding Antarctica. Hashed contours indicate statistically significant inter-model agreement on the sign of the trend at the α = 0.05 significance level. **b-e**, Regional analysis indicates statistically significant increasing trends in temporal autocorrelation in NHEX and TROP and a statistically significant decreasing trend in temporal autocorrelation in SHEX. While sea environments generally exhibit a greater degree of temporal autocorrelation than land, in NHEX autocorrelation is increasing at a greater rate at land locations as to overturn this relationship by the end of the 21st century.

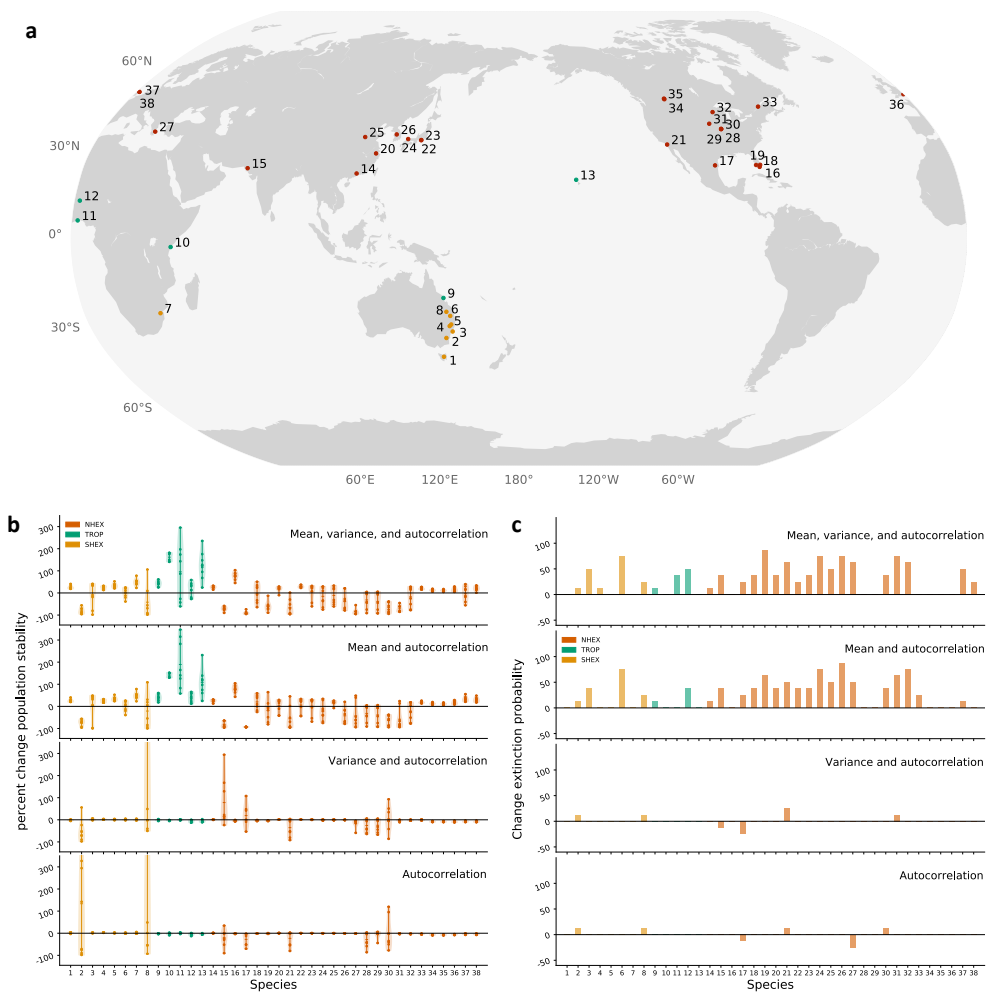

**Figure 4. Temperature has idiosyncratic effects on stability but increases extinction risk globally.**

**a**, Source locations of the terrestrial ectothermic invertebrate species, numbered 1 (southern-most latitude) to 38 (northern-most latitude). Species are color-coded according to latitudinal region (SHEX; 90°S to 23°S; orange, TROP; 23°S to 23°N; red, NHEX; 23°N to 90°N; green) **b**, Percent changes in population stability (mean÷standard deviation) between a historical reference period (1950-2000) and a future period (2050-2100) under multiple aspects of temperature change indicate greater risk to temperate than tropical species. Under a high emissions scenario, stability underwent a statistically significant increase for the plurality (16 of 38) of species and a statistically significant decrease for 10 species. Points in the violin plots represent the 8 climate model outputs. **c**, Extinction probability underwent a quasi-universal increase globally between the historical period (1950-2000) and a future period (2050-2100) under high emissions scenario changes in temperature.